\documentclass[aps,prd,twocolumn,showkeys,nofootinbib]{revtex4-1}
\usepackage[utf8]{inputenc}
\usepackage{amsmath}
\usepackage{amsfonts}
\usepackage{amssymb}
\usepackage{hyperref}
\usepackage{color}
\usepackage{graphicx}
\usepackage{epstopdf}
\usepackage{physics}
\usepackage{verbatim}
\usepackage{amsmath}
\usepackage{amsfonts}
\usepackage{multirow}
\usepackage{array,mathtools,booktabs}

\newcolumntype{C}{>{$}c<{$}}
\usepackage{cancel}
\AtBeginDocument{
\heavyrulewidth=.08em
\lightrulewidth=.05em
\cmidrulewidth=.03em
\belowrulesep=.65ex
\belowbottomsep=0pt
\aboverulesep=.4ex
\abovetopsep=0pt
\cmidrulesep=\doublerulesep
\cmidrulekern=.5em
\defaultaddspace=.5em
}
\usepackage{braket}
\usepackage{hyperref}
\hypersetup{
    colorlinks=true,
    linkcolor=blue,
    urlcolor=cyan,
    citecolor=cyan,}

\usepackage{fancyvrb}
\usepackage{subfig}
\numberwithin{equation}{section}

\renewcommand{\dd}{\mathrm{d}
}

\usepackage{siunitx}

\definecolor{klgreen}{rgb}{0.0, 0.5, 0.0}

\begin{document}

\title{Is the Chiral Magnetic Effect fast enough?}
\author{Jewel K. Ghosh$^{1,2}$}
\email{jewel.ghosh@iub.edu.bd}
\author{Sebastian Grieninger$^{3,4}$}
\email{sebastian.grieninger@gmail.com}
\author{Karl Landsteiner$^3$}
\email{karl.landsteiner@csic.es}
\author{Sergio Morales-Tejera$^{3,4}$}
\email{sergio.moralest@uam.es}
\affiliation{$^1$ Independent University Bangladesh (IUB), Plot 16, Block B, Aftabuddin Ahmed Road, Bashundhara R/A, Dhaka, Bangladesh\\ $2$ International Centre for Theoretical Sciences (ICTS-TIFR),
Tata Institute of Fundamental Research, Shivakote, Hesaraghatta, Bangalore 560089, India  \\$^3$Instituto de F\'isica Te\'orica UAM/CSIC, c/Nicol\'as Cabrera 13-15, Universidad Aut\'onoma de Madrid, Cantoblanco, 28049 Madrid, Spain\\$^4$Departamento de F\'isica Te\'orica, Universidad Aut{\'o}noma de Madrid, Campus de Cantoblanco, 28049 Madrid, Spain}

\preprint{IFT-UAM/CSIC-21-52}
\date{\today}

\captionsetup{justification=justified,singlelinecheck=false}

\begin{abstract}
It depends: While we find within holography that the lifetime of the magnetic field for collider energies like the ones achieved at RHIC is long enough to build up the chiral magnetic current, the lifetime of the magnetic field at LHC seems to be too short.

We study the real time evolution of the chiral magnetic effect out-of-equilibrium in
strongly coupled holographic gauge theories. 
We consider the backreaction of the magnetic field onto the geometry and monitor pressure and chiral magnetic current. Our findings show that generically at small magnetic field the pressure builds up faster than the chiral magnetic current whereas at strong magnetic field the opposite is true. At large charge we also find that equilibration is delayed significantly due to long lived oscillations. We also match the parameters of our model to QCD parameters and draw lessons of possible relevance to the realization of the chiral magnetic effect in heavy ion collisions. In particular, we find an equilibration time of about $\sim0.35$ fm/c in presence of the chiral anomaly for plasma temperatures of order $T\sim300-400$ M\si{\eV}. 
\end{abstract}

\maketitle

\newpage
\tableofcontents

\section{Introduction}
The chiral magnetic effect (CME) is the generation of an electric current in a chirally imbalanced medium by an applied magnetic field \cite{Vilenkin:1980fu,Fukushima:2008xe}.\footnote{For reviews see \cite{Kharzeev:2013ffa,Landsteiner:2016led,Kharzeev:2020jxw}.} Formally it is described by
\begin{equation}\label{eq:cmeeq}
\vec J = 8 c \mu_5 \vec B\,,
\end{equation}
where $c$ is the coefficient of the axial anomaly and $\mu_5$ the axial chemical potential. 

While this formula is rather straightforward to derive its interpretation is subtle in many respects. 
The fact that it depends on a chemical potential and the magnetic field implies that it can be derived in equilibrium quantum
field theory. This is indeed the case but the axial chemical potential is a thermodynamic variable conjugate to an
anomalous charge. The axial current is not conserved in the quantum theory but obeys the anomalous non-conservation law \cite{Adler:1969gk,Bell:1969ts}\footnote{In this work, we will not consider effects of the gravitational contribution to the axial anomaly \cite{Kimura:1969wj}.}
\begin{equation}\label{eq:anomaly}
\partial_\mu J^\mu_5 = c \epsilon^{\mu\nu\rho\lambda} F_{\mu\nu} F_{\rho\lambda}\,.
\end{equation}
Since a chemical potential should in principle be introduced only for conserved charges the question of the precise 
meaning of the axial chemical potential arises. An additional issue is that rather general arguments suggest that in equilibrium the electric current has to vanish identically \cite{Yamamoto:2015fxa}.  

This issue is well understood by now. The CME current indeed vanishes in equilibrium if one introduces the axial chemical potential as a background value of an axial gauge field $A_0^5=\mu_5$. The axial gauge field gives rise to an additional contribution to the CME that cancels the CME current \cite{Gynther:2010ed} (see also \cite{Rubakov:2010qi}). On the other hand, one can introduce the
chemical potential as a property of the initial state and then consider the time evolution generated by the Hamiltonian $H$
in which no axial gauge field is present \cite{Landsteiner:2012kd}. In that case eq. (\ref{eq:cmeeq}) does in fact apply. However, this already represents a
certain amount of non-equilibrium physics since the axial charge (or chemical potential) has to be induced by some means in the initial state. 

From these considerations it becomes clear that the physical realization of the CME demands for a certain amount of non-equilibrium physics. 
Hydrodynamics only assumes local thermal equilibrium and thus is always intrinsically a non-equilibrium theory too. 
In fact, it was shown that the chiral transport can be derived as a consequence of hydrodynamics with triangle anomalies \cite{Son:2009tf}. 

If one does not assume local thermal equilibrium but initiates the physical system in a generic non-equilibrium state the question arises how fast it evolves towards equilibrium. Since the CME current is formally expressed by a chemical potential (an equilibrium quantity) the question of how the CME is realized far from equilibrium is particularly interesting.  
Answering this question is particularly important in view of a possible realization of the CME in heavy ion collisions. The magnetic field in heavy ion collisions is only present in the initial stages and decays quickly whereas the hot QCD matter has a short but non-vanishing equilibration time. The question naturally arising is if the CME builds up fast enough to be measurable before the magnetic field decays. We want to address this question in this work by modelling the strongly coupled quark gluon plasma by means of a holographic model. 

The gauge/gravity duality or holographic duality provided many important insights into this type of questions. Starting with the key insight of the small value of the shear viscosity to entropy ratio $\eta/s = \frac{1}{4\pi}$ \cite{Kovtun:2004de}, holography has contributed many important results on hydrodynamics and transport in general. The modern way of understanding hydrodynamics is strongly influenced by it \cite{Baier:2007ix, Bhattacharyya:2008jc}. Many insights into anomaly induced transport phenomena have their origin in the holographic duality \cite{Banerjee:2008th,Erdmenger:2008rm,Landsteiner:2011iq,Ammon:2020rvg}. It also is an ideal tool to study far from equilibrium evolution of strongly coupled quantum systems \cite{Chesler:2008hg, Chesler:2013lia}. In the present work, we will therefore study the non-equilibrium behavior of the chiral magnetic effect in a holographic set-up. 

Previous studies of out-of-equilibrium chiral transport in holography are \cite{Lin:2013sga,Ammon:2016fru,Grieninger:2017jxz,Haack:2018ztx,Cartwright:2020qov}. 
A quantum simulation of the real time evolution of the CME has been presented in \cite{Kharzeev:2020kgc}.
In particular, the question of the timescale on which the CME becomes builds up has been the subject of \cite{Landsteiner:2017lwm,Fernandez-Pendas:2019rkh,Morales-Tejera:2020xuv}. In these works
a holographic approach was taken based on Vaidya type metrics. The advantage of this approach is that the background metric
can be computed analytically. The time evolution of chiral transport can then be studied numerically in linear response for
intrinsically small magnetic fields and beyond linear response in~\cite{Ammon:2016fru,Grieninger:2017jxz} in the probe limit. In contrast, the purpose of this work will be to study the full back-reaction of the magnetic field onto the geometry. 

The question of how fast the CME current builds up can be studied directly since it is possible to start with a far from
equilibrium state with large magnetic field in which no CME current is present. Another important observable that has been in the focus of holographic studies is the pressure~\cite{Chesler:2008hg}. The large magnetic field will of course induce a significant pressure anisotropy. The time evolution of the magnetic field induced pressure anisotropy in holography has been studied before in \cite{Fuini:2015hba} albeit without effects of the anomaly. In our case we chose to start with an initial state in which both, the CME current vanishes and the pressure anisotropy is not at its equilibrium value. We can then monitor the equilibration of both of our observables and study their time evolution for varying parameters. As parameters we chose the total energy $\epsilon$, the axial charge density $q_5$, the magnetic field $B$ and finally also the strength of the anomaly. The latter is represented in holography by the value of the Chern-Simons coupling $\alpha$. 

We will employ a bottom-up approach to construct our holographic model. The main motivation for choosing a bottom-up approach
is that the proper realization of the CME needs the notion of both an axial ($A)$ and vector $(V)$ like $U(1)$ symmetry. In holography,
we therefore need to introduce two bulk gauge fields and an appropriate Chern-Simons term representing the mixed $VAA$ anomaly.
It also allows to match the parameters of the model to QCD and we will do this later on by matching the holographic Chern-Simons coupling to the axial anomaly of three flavor QCD. 

The paper is structured as follows. In section \ref{sec:model} we present all the details of our holographic model. We set up the equations of motion and explain some of the salient features of the numerical methods.
Section \ref{sec:results} contains a scan through the parameter space. We present the time evolution of the CME current and the 
pressure anisotropy for varying magnetic field strength, varying Chern-Simons coupling and varying axial charge.
Of particular interest will be subsection \ref{subsec:matched} in which we match the Chern-Simons coupling to the value of
QCD with $N_f=3$. We summarize our findings and conclusions in section \ref{sec:conclusions}. Further details of the numerical methods will be presented in the appendix \ref{app:methods}. In appendix \ref{tab:appendix}, we provide results similar to section~\ref{subsec:matched} for a larger value of the axial charge density.

\section{The holographic model}\label{sec:model}
We  study a holographic quantum field theory with a $U(1)_A\times U(1)_V$ symmetry. The presence of two gauge fields $A^{\mu}$ and $V^{\mu}$ encodes the presence of the symmetry in the dual field theory. The field strengths of these fields are denoted as $F_5=d A$ and $F=dV$, respectively. The anomaly is implemented through a Chern--Simons term which is gauge invariant up to a total derivative. We work with the \textit{consistent} form of the anomaly. Combining these ingredients, the holographic model we consider is the following\footnote{In our notation, Greek letters denote the bulk coordinates and small Latin letters denote the boundary coordinates.}

\begin{align}
S=& \frac{1}{2\kappa^2}\int_{\mathcal{M}} d^5x\sqrt{-g}\left[R+\frac{12}{L^2}-\frac{1}{4}F^2-\frac{1}{4}F_{(5)}^2 \right. \nonumber \\
& \left. +\frac{\alpha}{3} \epsilon^{\mu\nu\rho\sigma\tau} A_\mu\left( 3 F_{\nu\rho}F_{\sigma\tau}+F^{(5)}_{\nu\rho}F^{(5)}_{\sigma\tau}\right) \right] \nonumber \\
& \qquad \qquad\qquad \qquad +S_{GHY}+S_{ct} \label{action} 
\end{align}
where $S_{GHY}$ is the Gibbons-Hawking-York boundary term to make the variational problem well defined, $L$ is the AdS radius, $\kappa^2$ is the Newton constant and $\alpha$ the Chern-Simons coupling. We also add appropriate counter-terms $S_{ct}$ to cancel the possible divergences \cite{deHaro:2000vlm,Balasubramanian:1999re,Emparan:1999pm}. The Levi-Civita tensor is defined as $\epsilon^{\mu \nu \rho \sigma \tau}=\epsilon(\mu \nu \rho \sigma \tau)/\sqrt{-g}\,$.  
The (consistent) anomaly is
\begin{eqnarray}
\delta_5 S = \frac{\alpha}{2\kappa^2} \int d^4x \lambda_5 \epsilon^{abcd} ( F_{ab} F_{cd} + \frac 1 3 F_{5,ab} F_{5,cd} )\,. 
\end{eqnarray}
The relative factor of $1/3$ reflects the Bose symmetry of the $U_A(1)^3$ anomaly, $\lambda_5$ is an axial gauge parameter.
\par
Varying the fields, we find the equations of motion. They are
\begin{align}
& \nabla_\nu F^{\nu\mu}+2\alpha \epsilon^{\mu\nu\rho\sigma\tau} F_{\nu\rho}F^{(5)}_{\sigma\tau}=0, \label{eom:Maxwell}\\
& \nabla_\nu F_{(5)}^{\nu\mu}+\alpha  \epsilon^{\mu\nu\rho\sigma\tau} \left( F_{\nu\rho}F_{\sigma\tau}+F_{\nu\rho}^{(5)}F_{\sigma\tau}^{(5)} \right)=0, \label{eom:axial}\\ 
& G_{\mu\nu}-\frac{6}{L^2} g_{\mu\nu}-\frac{1}{2} F_{\mu\rho}F_{\nu}^{\ \rho }
 +\frac{1}{8} F^2 g_{\mu\nu} -\frac{1}{2} F^{(5)}_{\mu\rho}F_{\nu}^{(5) \rho }\nonumber \\
 &+\frac{1}{8} F_{(5)}^{2}g_{\mu\nu}=0. \label{eom:Einstein}
\end{align}

We are interested in studying the vector current and pressure living in the dual field theory. Those can be extracted from the full bulk solution through the standard holographic prescription, i.e. varying the on-shell action with respect to the boundary value of the dual field appropriately subtracting the divergences.

In order to find the bulk solution we choose an ansatz for the metric and the gauge fields which preserves the symmetries present in the system.
We label the coordinates with $x^\mu=(u,v,x,y,z)$, where $u$ is the radial coordinate and $v$ the time coordinate.
In particular, we have translational invariance of the three spatial directions $(x,y,z)$. Hence, the metric fields can only depend on the radial $u$ and temporal $v$ coordinates. Besides, the presence of the magnetic field, which we assume to point in the $z$ direction, breaks the $SO(3)$ rotational invariance down to $SO(2)$. Consequently, we consider the metric ansatz to be 
\begin{eqnarray}
\label{eq:metricansantz}
& ds^2=-f(v,u) dv^2 - \frac{2L^2}{u^2} dvdu + \frac{2}{u^2} h(v,u)  dvdz \nonumber \\
 & + \Sigma(v,u)^2 \left[ e^{\xi(v,u)} (dx^2 + dy^2) + e^{-2\xi(v,u)} dz^2 \right],~~ \label{eq:ansatz}
\end{eqnarray}
which has been written in infalling Eddington-Finkelstein coordinates, with the boundary located at $u=0$. Note that $\xi$ parametrizes the anisotropy of the system. In order to recover asymptotic $AdS_5$, we demand 
\begin{equation}
\begin{split}
    & \lim_{u\to 0} f(v,u)=\frac{L^2}{u^2}\,, \hspace{0.8cm} \lim_{u\to 0}h(v,u)=0\,, \\ 
    &\lim_{u \to 0}\Sigma(v,u) = \frac{L}{u}\,, \hspace{0.8cm} \lim_{u \to 0}\xi(v,u)=  0\,. \label{eq:1.3}
\end{split}
\end{equation}
In this work, we consider a finite axial charge density and a external vector magnetic field $B$, which are the minimal ingredients required to generate the CME. We choose to work in the radial gauge $V_u=A_u=0\,$. 
Under these considerations, the only non-trivial component that is turned on for the axial gauge field is $A_v=-Q_5(v,u)$, whereas the vector gauge field contains $B$ and a non-trivial profile in its $z$-component.
Therefore, the ansatz for the gauge fields take the form

\begin{equation}
\begin{split}
&  V_{\mu}=(0,0,-y\,B/2,x\, B/2,V_z(v,u))\,,  \\
 & A_{\mu}=(-Q_5(v,u),0,0,0,0)\,. \label{gauge:A}
\end{split}
\end{equation}
 
The axial gauge field still has the gauge freedom to add a general function $g(v)$ to the temporal component without altering the physics. We fix this freedom by demanding that  $Q_5(v,u)$ vanishes at the boundary.\\
It turns out that such an ansatz allows to set the function $h(v,u)$ in \ref{eq:ansatz} to zero\footnote{The $h$ function would eventually account for the response of the stress tensor to the CME, yet only with $\mu_5$ and $B$ active the stress tensor receives no contribution.}, simplifying significantly the system of equations to solve.
Plugging the ansatz \ref{eq:ansatz} and \ref{gauge:A} into the equations of motion, setting $L=1$ and manipulating the expressions yields

\begin{align}
&    Q_5'=\frac{q_5}{u^2\Sigma^3}\left(1-8\alpha\frac{BV_z}{q_5} \right), \label{eq:Q5}\\
&    e^{2\xi}\Sigma (dV_z)'+\dfrac{1}{2}\left(e^{2\xi}\Sigma\right)'dV_z +\dfrac{1}{2}V_z'd\left(e^{2\xi}\Sigma\right)\nonumber \\
& -4\alpha B Q_5'=0\,, \label{eq:dV}\\
&    \frac{6 \Sigma'}{u \Sigma}+\frac{3 \Sigma''}{\Sigma}+\frac{ e^{2 \xi}}{2 \Sigma^2}V_z'^2+\frac{3}{2} \xi'^2=0\,,\label{eq:sigma} \\
&    3\frac{ d\Sigma'}{\Sigma}+\frac{6 d\Sigma \Sigma'}{\Sigma^2}-\frac{1}{4} u^2
   Q_5'^2+\frac{6}{u^2}-\frac{B^2 e^{-2 \xi}}{4 u^2 \Sigma^4}=0 \label{eq:dsigma}\\
&    (d\xi)'+ \frac{3}{2\Sigma}(\Sigma'd\xi+\xi'd\Sigma) -\frac{1}{6}\frac{B^2 e^{-2 \xi}}{ u^2 \Sigma^4} -\frac{e^{2 \xi}}{3
   \Sigma^2}V_z'dV_z= 0\,, \label{eq:dxi}\\
&      \dfrac{1}{2}(u^2f')'-\dfrac{3}{2}\xi'd\xi + 6 \dfrac{\Sigma'}{\Sigma} \dfrac{d\Sigma}{\Sigma} -\frac{5}{12}\frac{B^2 e^{-2 \xi}}{ u^2 \Sigma^4} + \dfrac{2}{u^2}\nonumber \\
&  - \frac{7}{12}u^2Q_5'^2-\frac{e^{2\xi}}{6\Sigma^2}V_z'dV_z=0\,,  \label{eq:f}\\
 &    dd\Sigma + \dfrac{1}{2}\Sigma d\xi^2 +\dfrac{1}{2}u^2f'd\Sigma + \frac{e^{2\xi}}{6\Sigma}dV_z^2 =0  \, \label{eq:cons},    
\end{align}

\noindent
where \ref{eq:Q5} has already been integrated once with integration constant $q_5$ which we eventually identify with the axial charge density. More specifically, setting $\mu=v,u$ in the axial Maxwell equations \eqref{eom:axial} we find
\begin{align}
   & \partial_u \left[ u^2 \Sigma^3 (v,u) \partial_u Q_5 (v,u)+8\alpha B V_z (v,u) \right]=0, \label{Maxwellu} \\
   & \partial_v \left[ u^2 \Sigma^3 (v,u) \partial_u Q_5 (v,u)+8\alpha B V_z (v,u) \right]=0. \label{Maxwellv}
\end{align}
Hence, the quantity $u^2 \Sigma^3 (v,u) \partial_u Q_5 (v,u)+8\alpha B V_z (v,u)$ is  independent of $u$ and $v$, and we identify it with the integration constant $q_5$. The prime and the dot denote radial and temporal derivatives respectively, whereas
\begin{equation}
 d=\partial_v-\dfrac{u^2 f}{2} \partial_u\,
\end{equation}
\noindent
is the derivative along infalling null geodesics. Its introduction is customary in this context and decouples some of the differential equations in a  nested structure.

We can use these equations of motion to find the near boundary expansions of the metric and gauge fields which read
\begin{align}
&  Q_5(v,u)=\frac{u^2}{2}q_5+\mathcal{O}(u^3)\,,\\
& V_z(v,u)=u^2\, V_{2}(v)+ \mathcal{O}(u^3)\,,\\
& \Sigma (v,u)=\frac{1}{u}+\lambda(v)+\mathcal{O}(u^5)\,, \\
& \xi(v,u)=u^4 \left( \xi_4(v)-\frac{B^2}{12} \,\log(u) \right)+\mathcal{O}(u^5) \,,\\
& f(v,u)=\left(\frac{1}{u}+\lambda(v)\right)^2 + u^2 \left( f_2+\frac{B^2}{6}\,\log(u) \right)  \nonumber \\
& -2 \Dot{\lambda}(v) +\mathcal{O}(u^3)\,. 
\end{align} \par 

\noindent
The function $\lambda(v)$ is a remnant of diffeomorphism symmetry and thus arbitrary. We follow \cite{Fuini:2015hba} and use $\lambda$ to keep the position of the apparent horizon of the black brane at a fixed radial position $u_h=1\,$ throughout the time evolution. The coefficient $f_2$ is related to the energy density of the black brane and the subleading coefficients $V_2(v)$ and $\xi_4(v)$ shall give us the vector current and the pressure anisotropy, respectively. In particular making use of the holographic prescription described above and substituting the asymptotic solution, we find  

\begin{equation}
\label{eq:QFTcurrent}
    2\kappa^2\left<J_z\right> = 2V_2(v)\,,\quad \quad 2\kappa^2\left<J_5^0\right> = q_5,
\end{equation}
\noindent
for the currents and 

\begin{equation}
\label{eq:stress}
    \begin{split}
        & \expval{T^v_{\ v}}=\frac{1}{4 \kappa^2}\left[ 6 f_2 -B^2 \textrm{log}(\mu L) \right]\,,\\
        & \expval{T^x_{\ x}}=\expval{T^y_{\ y}}=-\frac{1}{8\kappa^2}\left[B^2+4 f_2-16\xi_4(v)\right. \\
         &  \qquad \qquad \qquad \qquad\qquad \left. -2 B^2\textrm{log}(\mu L) \right]\,, \\
        & \expval{T^z_{\ z}}=-\frac{1}{4 \kappa^2} \left[ 2f_2+16\xi_4(v)+B^2 \textrm{log}(\mu L) \right]
    \end{split}
\end{equation}

\noindent
for the stress tensor. We have re-instated the AdS radius $L$ because in the regularization procedure a renormalization energy scale $\mu$ appears due to the fact that the magnetic field induces a trace anomaly, breaking conformal invariance at a microscopic level. All in all, the problem reduces to solving the full dynamics in the bulk, finding the subleading coefficients of $\xi$ and $V_z$ and substituting them into \ref{eq:QFTcurrent} and \ref{eq:stress}. The details about the numerical implementation of this strategy are summarized in the appendix \ref{app:methods}. Our numerical code is implemented in the programming language \textit{Julia}~\cite{bezanson2015julia}.

The thermodynamic properties of the system are specified by two quantities: temperature and the axial chemical potential. At late times, the system equilibrates and we shall label different solutions in terms of their thermodynamical variables of the final equilibrium state. Those will also be useful to compare with the QGP produced in heavy-ion collision experiments. In the dual gravity picture, this equilibration implies that the metric becomes stationary at late times. The temperature is formally that of the black brane once equilibrium is reached:

\begin{equation}
    T=\dfrac{1}{2\pi}\left.\left(-\dfrac{u^2}{2}\partial_u f(v,u) \right)\right|_{u=u_h \,,v\to\infty}\,.
\end{equation}
The chemical potential is computed as the temporal component of the gauge field at the boundary minus its value at the horizon, i.e. 
\begin{equation}\label{eq:defmu5}
    \mu_5 =\left. Q_5(v,u_h) - Q_5(v,0)\right|_{v\to \infty}\,.
\end{equation}  

We conclude this section by discussing the initial state of the dual quantum field theory. The asymptotic form of the metric ansatz \ref{eq:metricansantz} has been chosen so that it describes an infinite-volume non-expanding plasma. By construction the plasma has a charged distribution given by $q_5\,$, is immersed in a magnetic field of magnitude $B$ and has some energy density $\epsilon\,$. All three of them are considered to be homogeneous and constant in time in our model. Finally, we specify the initial conditions of the evolution by giving a starting profile to the fields $\xi$ and $V_z$. In particular, we choose them to be zero everywhere, which in turns means that the plasma starts with vanishing CME current and vanishing \textit{dynamical} pressure anisotropy, i.e. vanishing anisotropy generated by $\xi_4\,$. Notice that in \ref{eq:stress} the term proportional to $B$ does include anisotropy in the pressure from the beginning. That contribution is the same over the evolution and is referred to as $kinematic$ pressure anisotropy (see \cite{Fuini:2015hba} for further details). Hence, if there were no magnetic field we could interpret the initial conditions as being the equilibrium solution; yet since $B$ is present the system is out-of-equilibrium.

\section{Results of numerical simulations}\label{sec:results}
We scan the parameter space $(\alpha,q_5,B)$ and study the features displayed by the chiral magnetic effect through the vector current $\left<J\right>$\footnote{The vector current is parallel to the magnetic field, which we choose to be along the $z$-direction without loss of generality.} and the dynamical pressure anisotropy, which we denote $\xi_4$\footnote{Actually $\xi_4$ is proportional to the dynamical pressure anisotropy when evaluated at a scale $\mu=1/L\,$.}. The latter simply refers to the subleading coefficient of the metric field $\xi$, parametrizing the anisotropy. The transverse and longitudinal pressures can be trivially read off from \ref{eq:stress}.

There is a subtlety related to the definition of the energy density due to
the non-trivial renormalization scale dependence in (\ref{eq:stress})
\begin{equation}\label{eq:defenergy}
 \epsilon(\mu) = 2 \kappa^2 \langle T^{vv}\rangle\,.
\end{equation}
We refer the reader to the discussion in~\cite{Fuini:2015hba}
for details.
In order to assign a definite value to the energy-momentum tensor, we have to choose a renormalization scale $\mu$. 
For computational convenience, we choose $\mu=1/L$ throughout this section. The scale $1/L$ is, however, not a physical scale since it can be changed without changing the values of the physical observables on the boundary due to a scaling symmetry \cite{Fuini:2015hba}. 
The physically relevant scale is $\mu = \sqrt{B}$ which is also the natural choice. We will use this scale in section~\ref{subsec:matched}. Both choices are related through

\begin{equation}
    \dfrac{\epsilon_B}{B^2}=\dfrac{\epsilon_L}{B^2} + \dfrac{1}{4}  \textrm{log}(B L^2)\,,\label{eq:relebel}
\end{equation}

\noindent
where $\epsilon_L$ and $\epsilon_B$ refer to the energy density at scales $\mu=1/L$ and $\mu = \sqrt{B}$, respectively. 

 The parameter scan is performed at fixed energy density $\epsilon_L = 12\,$.\footnote{Working with a different value for $\epsilon_L$ seems to only modify the final equilibrium state for the pressure anisotropy but does not alter the relevant qualitative behavior like the build up time and the presence or absence of oscillations.} In section \ref{subsec:matched}, we match our model to QCD and give physically relevant values for the parameters.
 
\subsection{$B$-dependence}

\begin{figure}[h!]%
    \centering
    \subfloat{\includegraphics[scale=0.3]{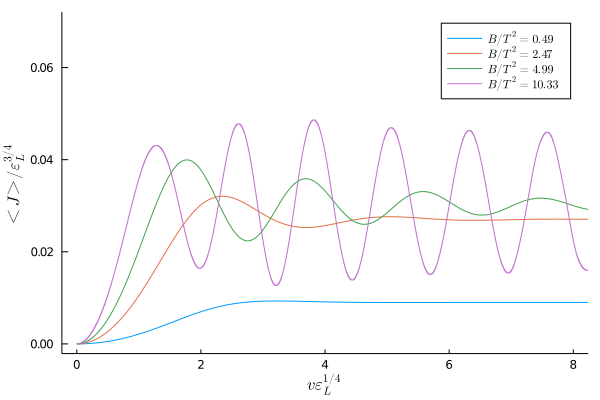}}%
    \qquad
    \subfloat{\includegraphics[scale=0.37]{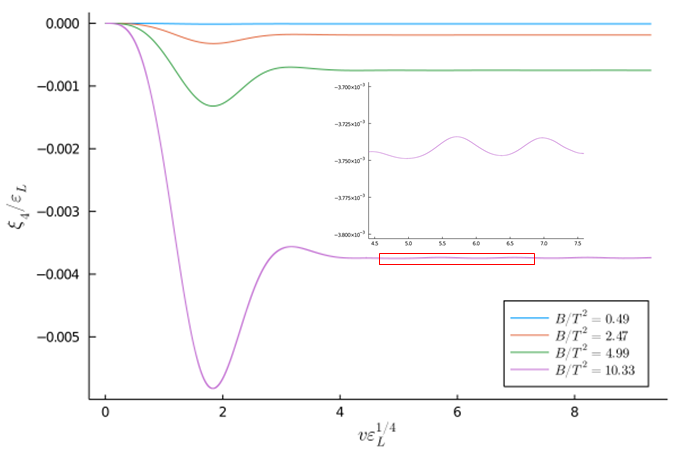}}
    \caption{Vector current (upper plot) and dynamical pressure anisotropy (lower plot) for fixed Chern-Simons coupling $\alpha=1.5$ and fixed small axial charge density $q_5=0.2\,$. The magnetic field $B$ is $\{0.1\,,0.5\,,1.0\,,2.0\}\,$. Even though $q_5$ is fixed, the dimensionless ratio of axial chemical potential to the temperature (in the final state) is different for each simulation. In particular, we find $\mu_5/T=\{0.11\,,0.06\,,0.04\,,0.02\}\,$ for the would be final equilibrium state.}
    \label{fig:1}
\end{figure}

\begin{figure}[h!]%
    \centering
    \subfloat{\includegraphics[scale=0.3]{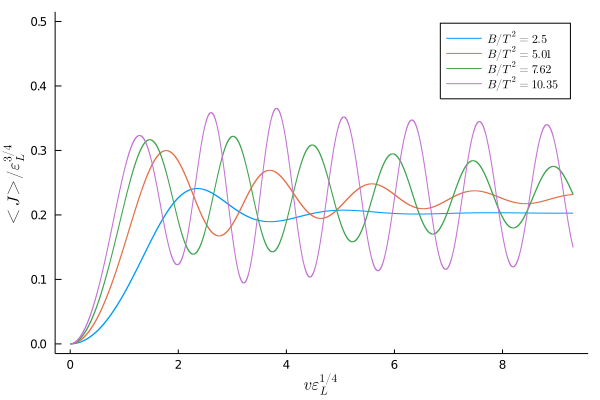}}%
    \qquad
    \subfloat{\includegraphics[scale=0.3]{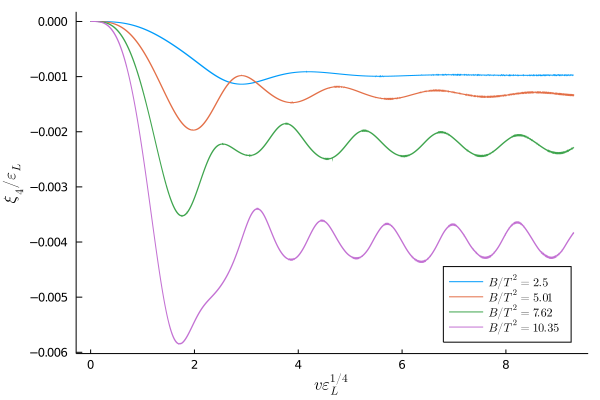}}
    \caption{Vector current (upper plot) and dynamical pressure anisotropy (lower plot) for fixed Chern-Simons coupling $\alpha=1.5$ and fixed large axial charge density $q_5=1.5\,$. The magnetic field $B$ is $\{0.5\,,1.0\,,1.5\,,2.0\}\,$. Even though $q_5$ is fixed, the dimensionless ratio $\mu_5/T$ (in the final state) is different for each simulation. In particular, we find $\mu_5/T=\{0.11\,,0.06\,,0.04\,,0.02\}\,$ for the would be final equilibrium state.}
    \label{fig:2}
\end{figure}

We first study both vector current and dynamical pressure anisotropy as we vary the vector magnetic field $B$. We keep the anomaly fixed, i.e. fixed Chern-Simons coupling $\alpha$, and consider two qualitatively different values of axial charge $q_5$. The results are shown in figures \ref{fig:1} and \ref{fig:2}. We choose to work with dimensionless variables: time, pressure and current are normalized to the energy density $\epsilon_L$, whereas we consider the dimensionless ratio of magnetic field $B$ to temperature squared. All thermodynamic quantities (chemical potential and temperature) refer to the final equilibrium state where they are well defined.

As we increase the magnetic field, we observe the appearance of oscillations in the current $\left<J\right>\,$. This is in agreement with the quasinormal modes computed in~\cite{Ammon:2016fru,Grieninger:2017jxz}, where they found that the imaginary part of the lowest QNM approaches the real axis for increasing the magnetic field and hence perturbations become long lived. As a consequence the equilibration time dramatically increases for increasing $B$. The oscillatory behavior of the current indicates that the time-evolution is dominated by the lowest QNM near the real axes. The final equilibrium value matches the equilibrium value for the CME, i.e. $2\kappa^2 \left<J\right>_{eq} = 8\alpha \mu_5 B\,$. Note that oscillatory behavior in the current indicates that we have not reached the final equilibrium state yet (and it may take a very long time to get there in the case of almost undamped oscillations). However, we verified that the axial chemical potential
\ref{eq:defmu5} read off from these states already closely matches the values of the would be equilibrium state.

It is also worth noting that the vector current builds up progressively faster with increasing magnetic field. 
We expect such a behavior for the following reason. At large magnetic field all fermions are in the lowest Landau level. The physics reduces effectively to the motion of the charged particles along the magnetic field lines and is thus effectively $1+1$ dimensional. In $1+1$ dimensions the following relation between the axial current and the vector current holds
\begin{equation}
    J^5_a = \epsilon_{ab} J^b\,,
\end{equation} (where the $a,b$ indexes are $v,z$). This is an operator relation and hence valid for matrix elements and expectation values. In contrast, the $3+1$ dimensional chiral magnetic effect depends on the (near-)equilibrium quantity $\mu_5$. Since for larger magnetic field the theory becomes more and more dominated by the effective $1+1$ dimensional dynamics, we expect the build up of the vector current to occur already in the non-equilibrium stages in order to fulfill the $1+1$ dimensional relation between the operators of axial charge and vector current.

Finally, increasing $q_5$ simply increases the absolute value of the final equilibrium state of the vector current. The effect in the pressure anisotropy is dramatically different. For large axial charge densities, we clearly observe oscillatory behavior in $\xi_4$. A closer look, however, reveals that the same type of oscillations are also present for small $q_5$ but their amplitude is significantly smaller and they could not be seen by eye in figure \ref{fig:1}, we have zoomed into one of the curves to clarify this statement.

\begin{figure}[h!]
    \centering
    \includegraphics[scale=0.4]{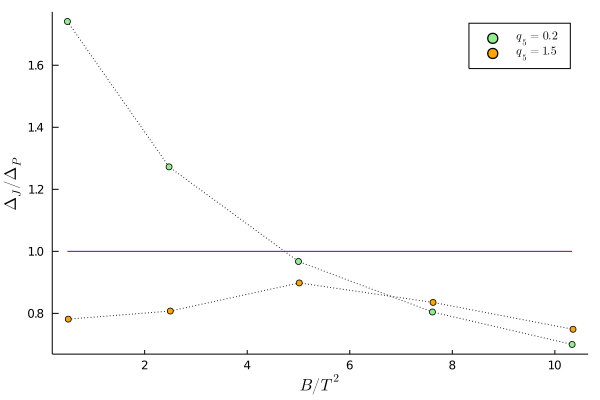}
    \caption{ Ratio of build up time for current and pressure anisotropy as a function of the magnetic field for small and high axial charge for fixed anomaly $\alpha = 1.5\,$.}
    \label{fig:3}
\end{figure}

We conclude this subsection with a discussion of the pressure anisotropy. To quantify our results we define the time where the first local extremum in the current and in the pressure anisotropy appears as \textit{build up time}. The build up time of the pressure anisotropy decreases slightly with increasing magnetic field. There then exists a crossover in the system as a whole as we vary $B$: for small $B$ the pressure anisotropy builds up faster than the current, whereas for large $B$ such behavior is reversed. As we increase axial charge the transition is lost and the vector current always builds up faster. We denote $\Delta_J$ and $\Delta_P$ for the vector current and pressure, respectively, and plot the ratio of both quantities in figure \ref{fig:3}.\\

%\newpage
\subsection{$\alpha$-dependence}

\begin{figure}[h!]%
    \centering
    \subfloat{\includegraphics[scale=0.3]{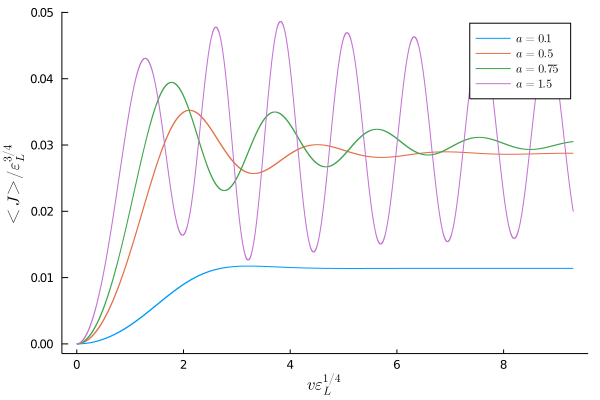}}%
    \qquad
    \subfloat{\includegraphics[scale=0.3]{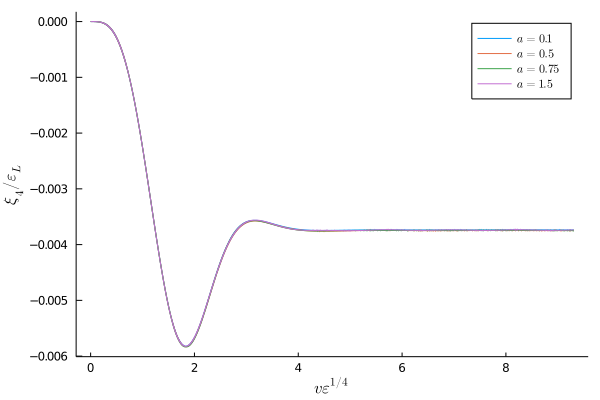}}
    \caption{Vector current (upper plot) and dynamical pressure anisotropy (lower plot) for fixed magnetic field $B=2$ and fixed small axial charge $q_5=0.2\,$. In dimensionless units, the simulations are for $B/T^2 = 10.34$ and a final equilibrium value of the axial chemical potential corresponding to $\mu_5/T=\{0.104\,,0.053\,,0.037\,,0.016\}$, respectively.}
    \label{fig:4}
\end{figure}

\begin{figure}[h!]%
    \centering
    \subfloat{\includegraphics[scale=0.3]{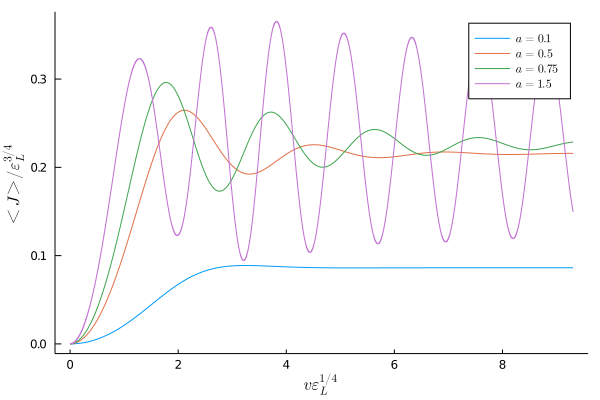}}%
    \qquad
    \subfloat{\includegraphics[scale=0.3]{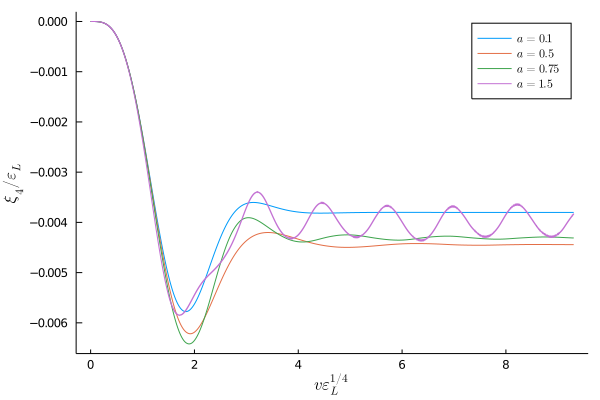}}
    \caption{Vector current (upper plot) and dynamical pressure anisotropy (lower plot) for fixed magnetic field $B=2$ and fixed large axial charge $q_5=1.5\,$. In dimensionless units, the simulations are for $B/T^2 = \{10.66\,,10.39\,,10.35\,,10.35\}\,$ and a final equilibrium value of the axial chemical potential corresponding to $\mu_5/T=\{0.802\,,0.396\,,0.278\,,0.119\}$, respectively.}
    \label{fig:5}
\end{figure}

In quantum field theory, the anomaly coefficient is fixed by the fermion spectrum. However, in the context of holography, the anomaly coefficient appears as a parameter of the holographic model that may be varied at will. We shall study its effect for two qualitatively different values of axial charge while keeping the magnetic field fixed at $B = 2\,$, or $B/\sqrt{\epsilon_L} = 0.58\,$ in dimensionless units. Results for small and high axial charge are shown in figures \ref{fig:4} and \ref{fig:5}, respectively.

\begin{figure}[h!]
    \centering
    \includegraphics[scale=0.4]{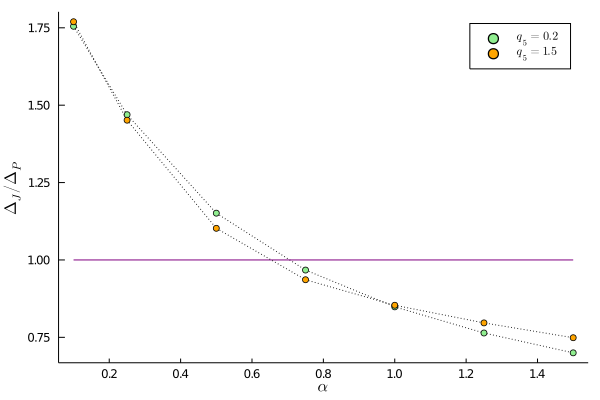}
    \caption{Ratio of build up time for the current and pressure anisotropy as a function of the Chern-Simons coupling $\alpha$ for small and high axial charge and fixed $B=2\,$.}
    \label{fig:6}
\end{figure}

The vector current builds up faster and develops oscillatory behavior as we increase $\alpha$ regardless of the magnitude of $q_5\,$. This is the same effect as we observed in the previous section. Hence, either increasing $\alpha$ or $B$ results qualitatively in analogous results. At small charge (see figure \ref{fig:4}), the evolution of the pressure seems to be governed solely by $B$ and is independent of $\alpha$. Zooming in on the tail of the curve shows that $\xi_4$ slightly depends on the anomaly coefficient $\alpha$, however, due to the small value of the axial charge the effect is negligible. The situation is clearer for larger axial charges $q_5$ (see figure \ref{fig:5}): increasing $\alpha$ yields again undamped oscillations, yet the build up time remains constant. In figure~\ref{fig:6} we show the ratio between the build up time of the vector current and pressure anisotropies as a function of the Chern-Simons coupling.  

\subsection{Matching to QCD}\label{subsec:matched}
\subsection*{Parameters}
In this section, we aim to provide simulations in the parameter range that is experimentally relevant for the Quark-Gluon plasma (QGP). We obtain estimates for the parameters in our model by matching to known QCD results, i.e. the entropy and the anomaly.\\

Under an axial gauge transformation, our action has the mixed anomaly $\mathcal{A}_{CS} = \frac{\alpha}{2 \kappa^2}$. In order to get an estimate for $\kappa$ we take the entropy of a black brane 

\begin{equation}
    s_{BH} = \dfrac{A}{4G_N} = \dfrac{4\pi^4T^3}{2 \kappa^2}\,.
\end{equation}

Now we want to match these expressions to the entropy of QCD at finite temperature and eventually to the anomaly. First, we need to fix how many flavors we take into account. The up and down quarks are light, whereas the strange quark has a mass of around
$95$ M\si{eV}. The cross over temperature of QCD is at around $175$ M\si{eV}. Therefore, we include the strange quark in our counting, i.e.
we match to three flavor QCD. The Stefan-Boltzmann value of the entropy density is

\begin{equation}
s_{SB} =  4 \left(\nu_b + \frac{7}{4} \nu_f\right) \frac{\pi^2 T^3}{90}\,.
\end{equation}

\noindent
where $\nu_b = 2(N_c^2-1)$ and $\nu_f = 2 N_c N_f$ with $N_c=3$ and $N_f=3$. Note that the Stefan-Boltzmann value of the entropy is reached only for asymptotically high temperatures. Typically, the
entropy at the temperatures of interest is lower. As a ballpark value, we take a factor of $3/4$ which is the one that arises
in the strongly coupled $\mathcal N=4$ SYM theory \cite{Gubser:1996de}. QCD lattice simulations also indicate a reduction by a factor of around $0.8$ at 
moderate temperatures (see e.g.  \cite{Borsanyi:2013bia}). Thus, we match our holographic model to QCD by

\begin{equation}
\frac{3 s_{SB}}{4} = s_{BH}\,,
\end{equation}

\noindent
from which we read off $\kappa^2 = (24 \pi^2)/19 \approx 12.5$. 

On the other hand, the axial anomaly in three flavor QCD is
\begin{equation}
\mathcal{A}_{QCD} = 2\,\frac{N_c}{32\pi^2}  \left(\frac 4 9 + \frac 1 9 +\frac 1 9\right) = \frac{1}{8\pi^2}\,,
\end{equation}
where the factor $2$ comes from the sum over right- and left-handed fermions and we sum over the squares
of the electric charges of up, down and strange quarks in the bracket. 
We can get the value for the Chern-Simons coupling by matching the anomaly $\mathcal{A}_{CS} = \mathcal{A}_{QCD}$ and this leads to

\begin{equation}
\alpha = \frac{6}{19} \approx 0.316\,.
\end{equation}

Let us finally discuss some physical considerations for the QGP. The strength of the temperature, magnetic field and chemical potential in typical nucleations of the QGP at RHIC and LHC are given in table~\ref{tab:parameters}.
\begin{table}[h!]
\begin{center}
\begin{tabular}{l c c}
\toprule
&RHIC	& LHC		\\ \addlinespace
\midrule
$T$		&$300\,\text{M}\si{\eV}$	&$1000\,\text{M}\si{\eV}$  \\
$\mu_5$		&$10\,\text{M}\si{\eV}$	&$10\,\text{M}\si{\eV}$  \\
$B$		&$m_\pi^2$	& $15\,m_\pi^2$\\
\bottomrule
\end{tabular}
\caption{
Parameters used in our simulations. For the temperature we take a lower value of roughly twice the critical temperature and a high value of roughly six times the critical temperature. The values for the magnetic field are taken from~\cite{Skokov:2009qp}. Estimates for the axial chemical potential are very uncertain and we take a small value of 10 M\si{eV} for both.
Due to the considerable uncertainties in the values of the parameters and also the lifetime of the magnetic field these should be viewed as ballpark values representative for RHIC and LHC physics.}
\label{tab:parameters}
\end{center}
\end{table}
\noindent

They provide us with two independent dimensionless quantities, which have to be adjusted in the numerical simulations with our two free parameters $(\epsilon_L, q_5)$. It turns out that fixing the dimensionless ratio $\epsilon_B/B^2$ gives a unique $B/T^2$, hence we work with $\epsilon_B$ and then compute the associated $\epsilon_L$ for the simulation through equation \eqref{eq:relebel}.  

In contrast to the previous sections, we show the full pressure anisotropy evaluated at the physical renormalization scale $\mu=\sqrt{B}$ in this section: 
\begin{equation}
  \delta P_i \equiv 2\kappa^2\,\frac{\Delta P_B}{B^2}=12\,\frac{\xi_4(v)}{B^2}+\frac 12\,\log(BL^2)-\frac{1}{4}.\label{eq:relpressure}
\end{equation}
Fixing $\epsilon_B/B^2$ in eq. \eqref{eq:relebel} does not fix $f_2$ and $B$ uniquely but rather gives us $B(f_2)$. This means that at fixed $\epsilon_B/B^2$ and vanishing initial dynamical anisotropy $\xi_4(0)=0$, we are confronted with a one parameter family of relative pressures of the initial state~\eqref{eq:relpressure} depending on the value of the magnetic field $B$ (for $L=1$). We shall exploit this feature to study equilibration of the pressure and current for several non-equivalent initial states by considering different values of $\delta P_i$.
\subsection*{Simulation}

\begin{figure}[h!]%
    \centering
    \subfloat{\includegraphics[scale=0.3]{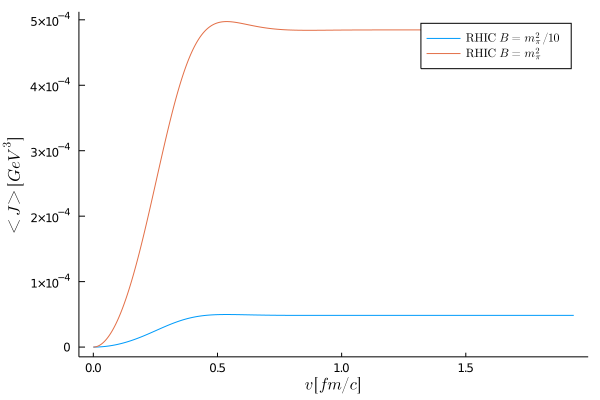}}%
    \qquad
    \subfloat{\includegraphics[scale=0.3]{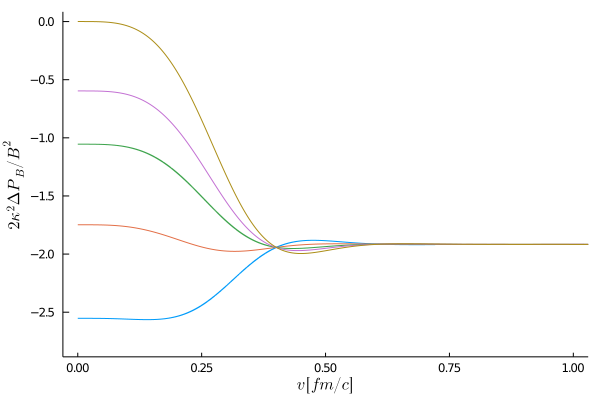}}
    \caption{Vector current (upper plot) and pressure anisotropy (lower plot) as a function of time for the physical parameter estimates for RHIC in table \ref{tab:parameters}, i.e. anomaly $\alpha\simeq 0.316$; for $m_\pi=140\,\text{M}\si{\eV}$. The pressure anisotropy is for $B=m_{\pi}^2$, the results for $B=0.1\,m_{\pi}^2$ are shown in figure \ref{fig:10}.}
    \label{fig:8}
    
\end{figure}

\begin{figure}[h!]%
    \centering
    \subfloat{\includegraphics[scale=0.33]{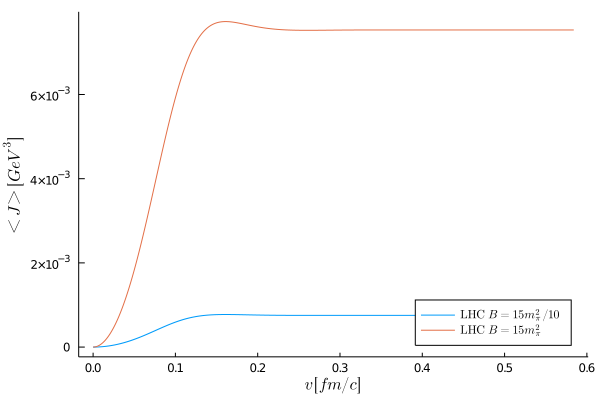}}%
    \qquad
    \subfloat{\includegraphics[scale=0.33]{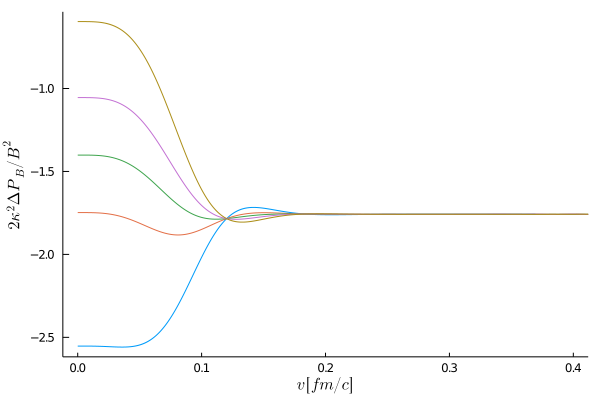}}
    \caption{Vector current (upper plot) and pressure anisotropy (lower plot) for the physical parameter estimates for LHC in table \ref{tab:parameters}, i.e. anomaly $\alpha\simeq 0.316\,$. The pressure anisotropy is for $B=15m_{\pi}^2$, the results for $B=1.5m_{\pi}^2$ are shown in figure \ref{fig:10}.}
    \label{fig:9}
    
\end{figure}

\begin{figure}[h!]%
    \centering
    \subfloat{\includegraphics[scale=0.3]{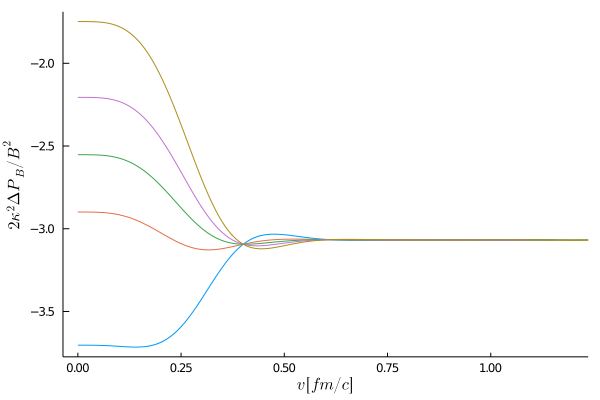}}%
    \qquad
    \subfloat{\includegraphics[scale=0.3]{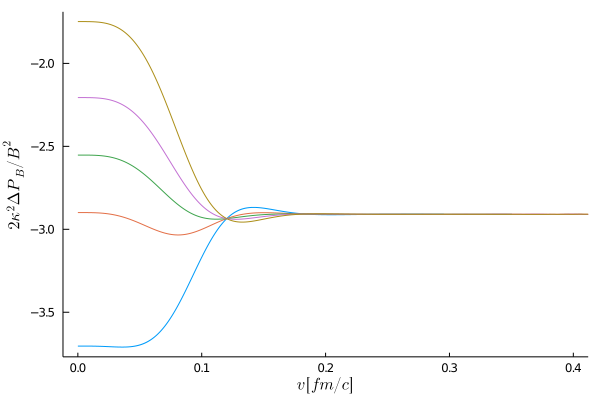}}
    \caption{Pressure anisotropy for RHIC (upper plots) and LHC (lower plot) with 10\% of the magnetic field compared to the pressures in figure~\ref{fig:8} and figure~\ref{fig:9}.}
    \label{fig:10}
    
\end{figure}

In figures \ref{fig:8}-\ref{fig:10}, we show the numerical results for the out-of-equilibrium CME with the physical parameters estimated in the previous sections. We fix our initial state by setting the dynamical pressure anisotropy to zero, i.e. $\xi_4(0)=0$, fix the ratio $\epsilon_B/B^2$ and the axial charge density $q_5$ so that we reach the temperature $T$ and the axial chemical potential $\mu_5$ indicated in table \ref{tab:parameters} as final equilibrium configuration. 

In figure~\ref{fig:8} and \ref{fig:9}, we present the results for the pressure and the current for the RHIC and LHC parameters, respectively.
Neither the vector current nor the pressure anisotropy shows oscillatory behavior. The former takes slightly more time to build up than the latter. On one hand, we observe that in the RHIC simulation in figure~\ref{fig:8}, the peak in the vector current is reached at $v_{peak}\sim 0.54 \,$fm/c. On the other hand, the pressure anisotropy reaches the peak at $v_{peak}\sim 0.48 \,$fm/c. 

We display the equilibration times for the simulation with the RHIC parameters in table~\ref{tab:RHICeq}. We use the definition of Chesler and Yaffe to label the equilibration time~\cite{Chesler:2008hg}, i.e. the time where the pressure anisotropy and the current, respectively, are within 10\% of their final value. As for the LHC in figure~\ref{fig:9}, we find the peak in the vector current at $v_{peak}\sim 0.14 \,$fm/c which is also the time where the pressure anisotropy reaches its peak. We tabulate the equilibration times for the simulation with the LHC parameters in table~\ref{tab:LHCeq}. Note that the equilibration times for the LHC parameters are about 1/3 shorter than in the RHIC case.

Another interesting feature is that the equilibration time for the pressure slightly depends on the initial state in a non-monotonous fashion. The change in tendency can be understood as a consequence of choosing an inital state with a pressure anisotropy that is either above or below the final equilibrium state. Actually, in the regime where the initial and final states do not differ more than $10\%$ the equilibration time prescribed above is rather ill-defined, because we could have a curve in whose pressure does not deviate much from the final value yielding $v_{eq}=0\,$. However, this problem does not arise for the parameters chosen in our simulations.

An estimate for the lifetime of the magnetic field has recently been given in \cite{Guo:2019joy} as $\tau_B \sim \frac{115\,\, \textrm{G}\si{eV}\textrm{fm/c}}{\sqrt{s}}$, where $\sqrt{s}$ is the energy of the collision. At RHIC and LHC the collisions take place at around $\sqrt{s}\simeq 200 \,\textrm{G}\si{eV}$ and $\sqrt{s}\simeq 5000 \,\textrm{G}\si{eV}$, respectively, which yield lifetimes of $\tau_B^\text{RHIC}\sim 0.6 \, \textrm{fm/c}$ and $\tau_B^\text{LHC}\sim 0.02 \, \textrm{fm/c}\,$. 
In this context the equilibration and build up times extracted from our simulations are of high significance.
It is clear from the equilibration times that for the RHIC parameter choice the current reaches its equilibrium value before the magnetic field vanishes. On the contrary, for the LHC parameter choice the magnetic field is short lived and is gone before the current could start to build up. Hence, we conclude that the chiral magnetic effect should only be observable at RHIC and not at LHC.

We notice that $\Delta P/B^2$ for fixed $\epsilon_B/B^2$ yields the same final equilibrium state independent of the value of $B$ as we expect. Even though the initial state is different for each run all curves cut at the same point during the evolution. The current is not influenced by the specific choice of $B$ as long as the dimensionless ratios stay constant.

In heavy-ion collisions, the magnetic field drops almost instantaneously from its peak value indicated in table~\ref{tab:parameters} where it stays for most of its remaining lifetime. Since we consider the magnetic field as static and the drop happens almost instantaneously, we did a second simulation for our parameter estimates with 10\% of the peak magnetic field~\ref{tab:parameters}. The corresponding results for the current are the blue curves in figure~\ref{fig:8} and and figure~\ref{fig:9} and the results for the pressure are depicted in figure~\ref{fig:10}. Even though the smaller magnetic field influences the overall absolute values of the observables, the equilibration times remain effectively unchanged which may be seen from the results tabulated in the the lower columns of table~\ref{tab:RHICeq} and ~\ref{tab:LHCeq}.

\begin{table}[h!]
\begin{center}
\begin{tabular}{l c c c c c}
\toprule
RHIC $B=m_\pi^2$ &&&&&  		\\ \addlinespace
\midrule
$\delta P_i$ &-2.55&-1.75&-1.05&-0.60&0.00  \\
$v^{\langle J\rangle}_\text{eq}$ \hspace{0.22cm}in [fm/c]		 &0.380&0.380&0.380&0.380&0.380  \\
$v^{\langle \Delta P\rangle}_\text{eq}$ in [fm/c]		 &0.383&0.418&0.334&0.344&0.350 \\\addlinespace
\toprule
RHIC $B=0.1m_\pi^2$ &&&&&  		\\ \addlinespace
\midrule
$\delta P_i$ &-3.70&-2.90&-2.55&-2.21&-1.75  \\
$v^{\langle J\rangle}_\text{eq}$ \hspace{0.22cm}in [fm/c]		 &0.380&0.380&0.380&0.380&0.380  \\
$v^{\langle \Delta P\rangle}_\text{eq}$ in [fm/c]		 &0.383&0.418&0.310&0.334&0.344 \\
\bottomrule
\end{tabular}
\caption{Equilibration times $v_\text{eq}$ for the RHIC simulation at $B=m_\pi^2$ and $B=0.1\,m_\pi^2$; $\delta P_i$ labels the different initial conditions for the pressure anisotropy~\eqref{eq:relpressure}.}
\label{tab:RHICeq}
\end{center}
\end{table}
\begin{table}[h!]
\begin{center}
\begin{tabular}{l c c c c c}
\toprule
LHC $B=15\,m_\pi^2$ &&&&&  		\\ \addlinespace
\midrule
$\delta P_i$ &-2.55&-1.75&-1.40 &-1.05&-0.60 \\
$v^{\langle J\rangle}_\text{eq}$ \hspace{0.22cm}in [fm/c]		 &0.114&0.114&0.114&0.114&0.114  \\
$v^{\langle \Delta P\rangle}_\text{eq}$ in [fm/c]		 &0.114&0.187&0.085&0.098&0.103 \\\addlinespace
\toprule
LHC $B=1.5\,m_\pi^2$ &&&&&  		\\ \addlinespace
\midrule
$\delta P_i$ &-3.70&-2.90&-2.55&-2.21&-1.75  \\
$v^{\langle J\rangle}_\text{eq}$ \hspace{0.22cm}in [fm/c]		 &0.114&0.114&0.114&0.114&0.114  \\
$v^{\langle \Delta P\rangle}_\text{eq}$ in [fm/c]		 &0.114&0.187&0.085&0.098&0.103 \\
\bottomrule
\end{tabular}
\caption{Equilibration times for the LHC simulation at $B=15\,m_\pi^2$ and $B=1.5\,m_\pi^2$; $\delta P_i$ labels the different initial conditions for the pressure anisotropy~\eqref{eq:relpressure}.}
\label{tab:LHCeq}
\end{center}
\end{table}

In the parameter estimates in table~\ref{tab:parameters}, the estimate for the axial chemical potential is the most uncertain one in the literature. To prove that our estimations for the build up and equilibration times of the current and the pressure are not influenced by choosing the particular value of $\mu_5=10\textrm{M}\si{eV}$, we provide analogous simulations at a ten times larger axial chemical potential of $\mu_5=100\textrm{M}\si{eV}$ in appendix~\ref{app:extrarun}. The time dependent current and the pressure anisotropy are depicted in figure~\ref{fig:11} and figure~\ref{fig:12}, respectively. Furthermore, we tabulated the equilibration and build up times in table~\ref{tab:appendix}. The bottom line is that our results for the build up times and thus the presence of the chiral magnetic effect in heavy ion collisions at RHIC and LHC remain qualitatively unchanged at the larger axial chemical potential.

\section{Conclusions}\label{sec:conclusions}
In this work, we investigated the out-of-equilibrium behavior of the chiral magnetic effect in the presence of strong external magnetic fields. We characterize how the chiral anomaly, the magnetic field and the axial charge density influence the non-equilibrium response of the chiral magnetic vector current and the pressure anisotropy and how they affect their equilibration and build up times.

To quantify the real time response of the vector current and the anisotropy, we performed a parameter scan. For a fixed strength of the chiral anomaly, we investigated the dependence of the response on the magnetic field $B$ for a small and large value of the axial charge density $q_5$. Increasing the magnetic field at fixed strength of $q_5$ eventually leads to long lived oscillations in the vector current which send the equilibration time to infinity. This is in agreement with the QNM results for our system obtained in~\cite{Ammon:2016fru,Grieninger:2017jxz}. Furthermore, the build up time of the vector current gets progressively smaller for increasing the magnetic field. Both effects might be rooted in the presence of Landau levels in our system. For large magnetic fields, the system is effectively 1+1 dimensional and the physics is totally dictated by the lowest Landau level. Keeping the magnetic field constant while increasing the axial charge density simply increases the final value of the current. The build up time for the anisotropy also decreases for increasing the magnetic field even though the effect is small. However, increasing the axial charge density dramatically affects the pressure anisotropy since it induces long lived oscillations which appear to be absent in the setup without chiral anomalies~\cite{Fuini:2015hba}. Indeed, we show explicitly that the anomaly coefficient has to be sufficiently large in order to observe these long lived oscillations.

Interestingly, we observe a crossover in the build up times of the vector current and the anisotropy at small axial charge. For small magnetic fields, the pressure anisotropy builds up faster while at large magnetic field the roles are reversed. For large axial charges, the vector current always builds up faster than the anisotropy independent of the magnetic field.

Finally, we aim to provide insights on the build up time of the chiral magnetic current in heavy ion collision experiments at RHIC and LHC. Within our setup, the build up time of the chiral magnetic effect is smaller than the lifetime of the magnetic field and should thus be an observable in heavy ion collisions at RHIC~\cite{Shi:2019wzi}. However, the lifetime of the magnetic field at LHC seems to be so short that the magnetic field already drops to zero before the chiral magnetic current can build up in a meaningful way. Furthermore, we find that in both cases, the build up time of the chiral-magnetic current is approximately as fast as the build up time of the pressure anisotropy. Interestingly, in the RHIC case we find in presence of the chiral anomaly a shorter equilibration time of $\sim0.35$ fm/c (for an initial state with $\delta P_i(0)=0$) compared to the result of Chesler and Yaffe which estimates the equilibration time as $\sim0.5$ fm/c~\cite{Chesler:2008hg}. This is in agreement with the equilibration time estimate of ~$\sim 0.3$ fm/c for plasma temperatures of $T\sim300-400$ MeV~\cite{Heinz:2004pj}. Note that the build up time of the chiral magnetic current is with $\sim 0.38$ fm/c in the same parameter range.
The parameter estimate for the axial chemical potential seems to be the most uncertain one in the literature. To prove that our results do not rely on the given parameter estimate of $\mu_5=10\textrm{M}\si{eV}$, we verified that our results for the build up and equilibration times remain qualitatively unchanged for a ten times larger axial chemical potential.

In the future, it would be interesting to generalize our setup to more realistic case of asymmetric shockwave collisions as initiated in \cite{Muller:2020ziz,Waeber:2019nqd}. It would also be very interesting to consider time dependent, dynamical magnetic fields as they are present in heavy-ion collisions. Finally, it would be interesting to understand how the quantum critical point investigated in~\cite{DHoker:2010zpp,DHoker:2012rlj,Ammon:2016szz} influences the non-equilibrium dynamics of the system. We leave these questions open for future investigations.

\acknowledgments{SG is supported by the `Atracci\'on de Talento' program (2017-T1/TIC-5258, Comunidad de Madrid) and through the grants SEV-2016-0597 and PGC2018-095976-B-C21.  SMT is supported by an FPI-UAM predoctoral fellowship.}
%\newpage
\appendix
\section{Numerical methods}\label{app:methods}
In this appendix, we explain the numerical methods used throughout this work.
We use the so-called characteristic formulation of Bondi and Sachs\footnote{Note that there is a second approach from numerical relativity established in \cite{Heller:2012je}.} established in holographic setups by~\cite{Chesler:2008hg,Chesler:2009cy,Chesler:2010bi,Chesler:2013lia,Baggioli:2019mck}. The big advantage of this approach is that a set of coupled partial differential equations decouple in a nested structure of ordinary differential equations in which the equations may be solved successively. In terms of the characteristic formulation, we solve the ordinary differential equations on a given time-slice by means of pseudo-spectral methods. 
The main idea of pseudo-spectral methods (for an introduction see~\cite{Boyd1989ChebyshevAF}; here we follow~\cite{Grieninger:2020wsb,Baggioli:2019abx,Baggioli:2020edn}) is to expand the solution $u(x)=\sum_{n=0}^\infty\, c_n\phi_n(x)$ to the differential equation in a basis $\{\phi_n(x)\}$ and approximate the exact solution $u(x)$ by a finite number $N$ of basis polynomials $\phi_n(x)$
\begin{equation}
    u(x)\approx u_N(x)=\sum\limits_{n=0}^N\,c_n\,\phi_n(x).\label{eq:spectralrep}
\end{equation}
As basis functions we choose Chebychev functions
\begin{equation}
    T_k(x)=\cos(k\,\arccos(x)).
\end{equation}
We can re-write first and second derivative by using the derivatives of the basis functions i.e. $\phi'_m(x)=\sum_{n=0}^N\hat D_{mn}\phi_n(x), \ \phi''_m(x)=\sum_{n,l=0}^N\hat D_{mn}\hat D_{nl}\,\phi_l(x)$.  With the differentiation matrices, we can rewrite derivatives so that they act on the coefficients, for example
\begin{align}
    &u'(x) \approx\!\sum\limits_{n=0}^N c_n\,\phi_j'(x)=\!\sum\limits_{n,m=0}^Nc_n\,\hat D_{nm}\phi_m(x)\!=\!\sum\limits_{n=0}^Nc_n'\,\phi_n(x).
\end{align}
To discretize the differential equations in the radial direction, we use a Chebychev-Lobatto grid with $N$ gridpoints given by 
\begin{equation}
    x_i=\cos\varphi_i=\cos\frac{\pi \,i}{N}.
\end{equation}

We may solve the equations of motion for the axial gauge field by introducing
\begin{equation}
q_5\equiv8 \alpha\, B\, V(v,u)+u^2 Q_5'(v,u) \Sigma(v,u)^3
\end{equation}
and simplify it further by introducing $\tilde q_5(v, u, \alpha, B) = q_5 - 8 \alpha\, B\, V(v, u)$. 
As explained in~\cite{Chesler:2013lia}, the condition for fixing the apparent horizon to $u_h=1$ on the initial time slice reads $\dd \Sigma(u_h)=0.
$
We can keep the apparent horizon fixed at $u_h=1$ by imposing that the time derivative of the aforementioned equation vanishes. By using the equations of motion we find that we can implement this as a boundary condition on the blackening factor $f$ at the horizon, i.e.
\begin{align}
&2\,f(v,1) \left(B^2 \Sigma(v,1)^2 e^{-2\, \xi(v,1)}+\tilde q_5^2-24 \,\Sigma(v,1)^6\right)\label{eq:horizon}\\&\!\!\!-2 \,\Sigma(v,1)^4\! \left(\dd V(v,1)^2\, e^{2\, \xi(v,1)}\!+3\, \dd \xi(v,1)^2\, \Sigma(v,1)^2\right)\!\!=0\nonumber
\end{align}
We start with an initial profile for $V$ and $\xi$. The initial data also contains the energy density $\epsilon=-3\,f_2$, the axial charge density $q_5$, the Chern-Simons coupling $\alpha$ and the magnetic field $B$. On a given time slice we solve the set of equations in the following order: eq. \eqref{eq:sigma} for $\Sigma$, eq. \eqref{eq:dsigma} for $\dd\Sigma$, \eqref{eq:dV} and eq. \eqref{eq:dxi} for $\dd V$ and $\dd\xi$, eq. \eqref{eq:f} for the blackening factor $f$. Eq. \eqref{eq:cons} functions as a constraint. Every time step, we can extract $\dot\lambda$ by reading of $\dot\lambda=-f_s(u=0)/2$. We impose the constraint equation \eqref{eq:cons} in terms of the horizon boundary condition for $f$~\eqref{eq:horizon} and at the conformal boundary in terms of $\dd \Sigma_s(u=0)=0$. Finally, to evolve in time we use a forth order Runge-Kutta with an appropriately small timestep.

\begin{figure}[h]
    \centering
    \subfloat{\includegraphics[scale=0.3]{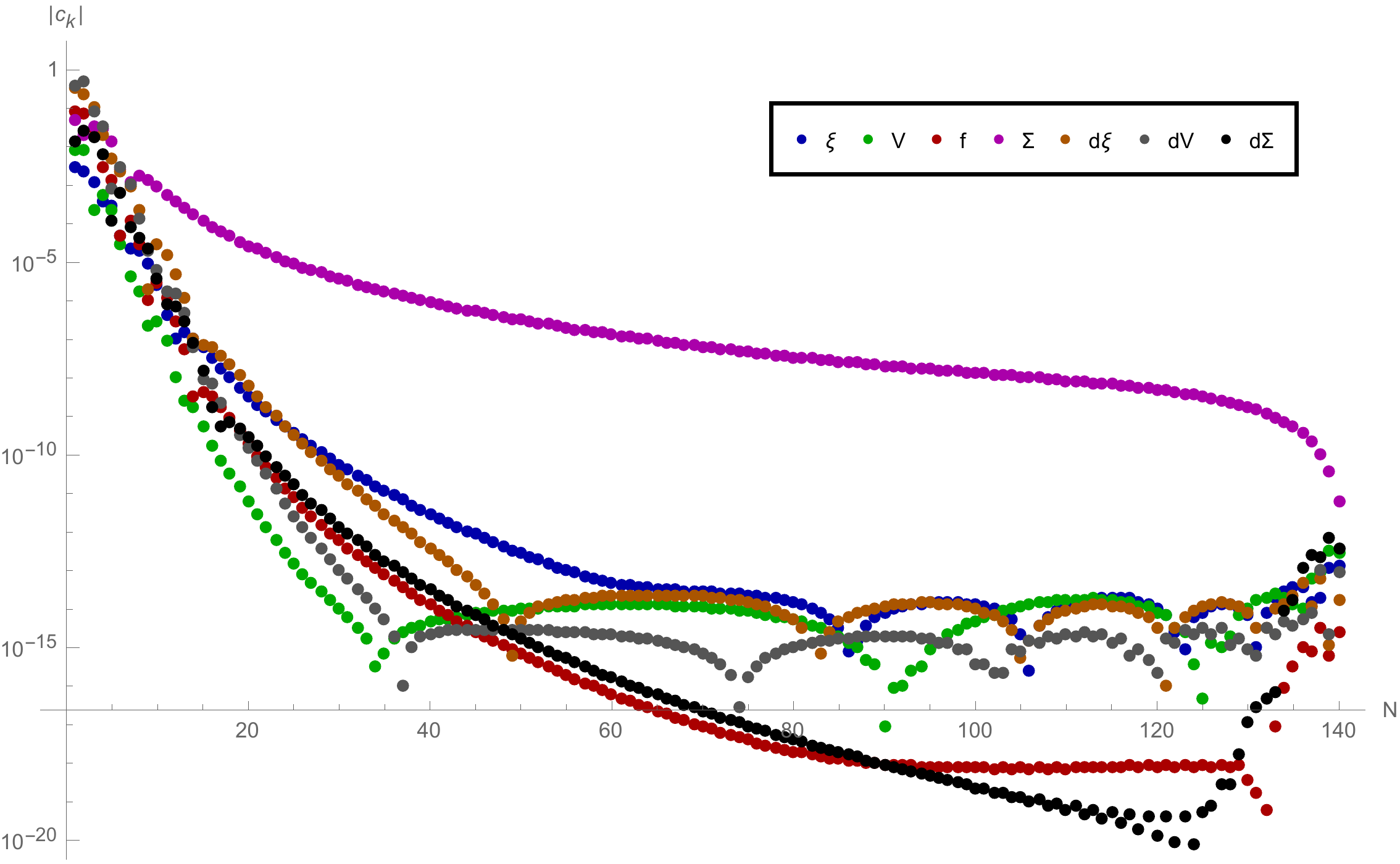}}%
    \caption{Chebychev coefficients for a large magnetic field $B/\sqrt{\epsilon}=2.31,\, q_5/\epsilon^{3/4}=0.31,\,\alpha=6/19$.}
    \label{fig:cheby}
\end{figure}

To improve the convergence we subtract the logarithmic terms up to appropriated order and work with redefined functions which are given by%\newpage
\begin{align*}
\Sigma(v,u)=&\frac 1u+\lambda(v)+u^{5}\Sigma_s(v,u)\\
\dd \Sigma(v,u)=&\frac 12\,\Sigma(v,u)^2+\dd \Sigma_s(v,u)+\epsilon\,\left(\frac1u+\lambda(v)\right)^{-2}\\&-\frac{B^2}{12} \log\!\left(\frac 1u+\lambda(v)\right)\,\left(\frac1u+\lambda(v)\right)^{-2}\\
\xi(v,u)=&\left(\frac1u+\lambda(v)\right)^{-3}\, \xi_s(v,u)\\&-\frac{B^2}{12} \log\!\left(\frac 1u+\lambda(v)\right)\,\left(\frac1u+\lambda(v)\right)^{-4}\\
\dd \xi(v,u)=&\left(\frac1u+\lambda(v)\right)^{-2}\, \dd \xi_s(v,u)\\&-\frac{B^2}{6} \log\left(\frac 1u+\lambda(v)\right)\,\left(\frac1u+\lambda(v)\right)^{-3}\\
f(v,u)=&\left(\frac1u+\lambda(v)\right)^2\!\!+f_s(v,u)+2 \epsilon\,\left(\frac1u+\lambda(v)\right)^{-2}\\&-\frac{B^2}{6} \log\!\left(\frac 1u+\lambda(v)\right)\,\left(\frac1u+\lambda(v)\right)^{-2}
\\
V(v,u)&=\left(\frac1u+\lambda(v)\right)^{-1} \,V_s(v,u)\\
\dd V(v,u)&=\dd V_s(v,u).\end{align*}
We monitor the accuracy of our numerical algorithm by different methods. Firstly, we check the constraint equation throughout the time evolution and monitor how accurately the apparent horizon stays at one. Secondly, we compared our solution to a solution with a larger number of gridpoints and checked that it does not change significantly. Lastly, we checked the Chebychev coefficients of our numerical solution, as presented in figure~\ref{fig:cheby} for a given time and ensured that the coefficients drop to the required precision.

\section{Simulations for $\mu_5=100$MeV.}\label{app:extrarun}
\begin{figure}[h!]%
    \centering
    \subfloat{\includegraphics[scale=0.33]{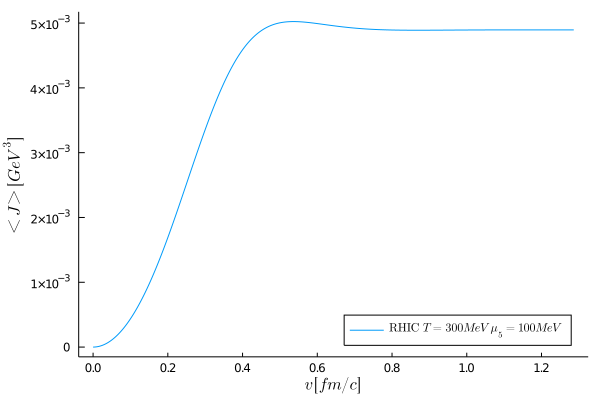}}%
    \qquad
    \subfloat{\includegraphics[scale=0.33]{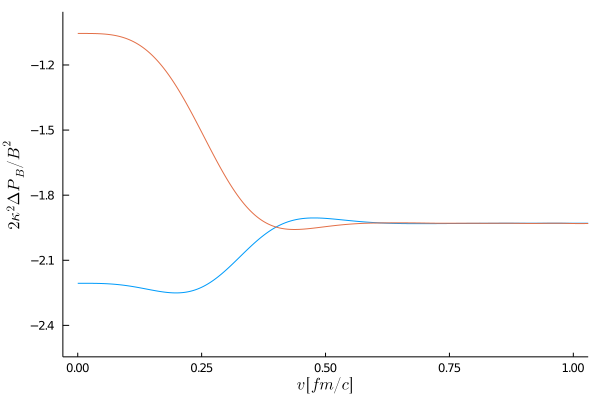}}
    \caption{Vector current (upper plot) and pressure anisotropy (lower plot) for the physical parameter estimates for RHIC at chemical potential $\mu_5=100\,\textrm{M}\si{eV}$ (otherwise with the values from table \ref{tab:parameters}) and anomaly $\alpha\simeq 0.316\,$.}
    \label{fig:11}
\end{figure}

\begin{figure}[h!]%
    \centering
    \subfloat{\includegraphics[scale=0.33]{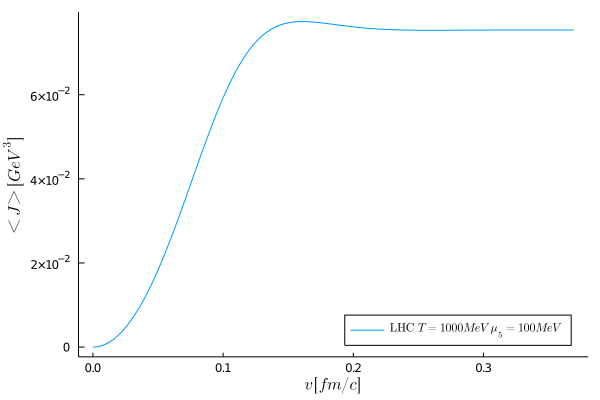}}%
    \qquad
    \subfloat{\includegraphics[scale=0.33]{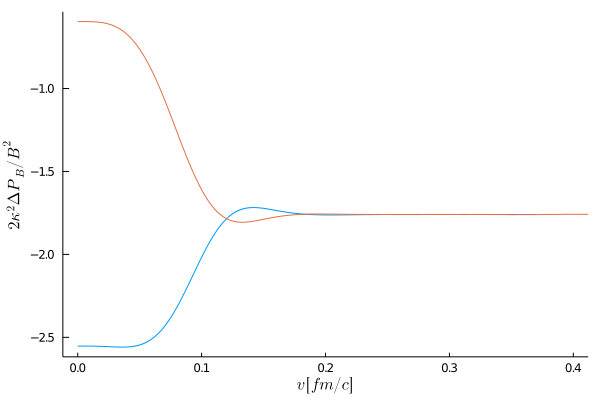}}
    \caption{Vector current (upper plot) and pressure anisotropy (lower plot) for the physical parameter estimates for LHC at chemical potential $\mu_5=100\,\textrm{M}\si{eV}$ (otherwise with the values from table \ref{tab:parameters}) and anomaly $\alpha\simeq 0.316\,$.}
    \label{fig:12}
\end{figure}

In this appendix, we provide simulations with a ten times larger chemical potential compared to the parameter estimates given in table~\ref{tab:parameters}. We keep all the other parameters the same and figure~\ref{fig:11} has to be compared with figure~\ref{fig:8} (for RHIC at $B=m_\pi^2$) and figure~\ref{fig:12} with figure~\ref{fig:9} (for LHC at $B=15\,m_\pi^2$). We tabulate the equilibration and build up times for these simulations in table~\ref{tab:appendix}. 

\begin{table}[h!]
\begin{center}
\begin{tabular}{l c c c c}
\toprule
\multicolumn{1}{l}{$\mu_5=100\textrm{M}\si{eV}$}&\multicolumn{2}{c}{RHIC ($B=m_\pi^2$)} &
\multicolumn{2}{c}{LHC ($B=15\,m_\pi^2$)} \\
\cmidrule(lr){2-3}
\cmidrule(lr){4-5}
$\delta P_i$&
\multicolumn{1}{c}{-2.21}&
\multicolumn{1}{c}{-1.05}&
\multicolumn{1}{c}{-2.55}&
\multicolumn{1}{c}{-0.60}\\
\addlinespace
\midrule
$v^{\langle J\rangle}_\text{eq}$ \hspace{0.22cm}in [fm/c]		 &0.380&0.380&0.114&0.114  \\
$v^{\langle \Delta P\rangle}_\text{eq}$ in [fm/c]		 &0.393&0.336&0.113&0.102 \\\addlinespace
\midrule
$v_\text{peak}^{\langle J\rangle}$ \hspace{0.185cm}in [fm/c]		 &0.537 &0.537 &0.161&0.161  \\
$v_\text{peak}^{\langle \Delta P\rangle}$ \hspace{0.08cm}in [fm/c]		 &0.477&0.437&0.142&0.133 \\\addlinespace
\bottomrule
\end{tabular}
\caption{Equilibration and build up times for the RHIC and LHC simulation at ten times the axial charge density.}
\label{tab:appendix}
\end{center}
\end{table}

\bibliography{HoloCME}

\end{document}